\documentclass[aps,prl,twocolumn,showpacs,superscriptaddress,groupedaddress]{revtex4}  
\usepackage{graphicx}  
\usepackage{dcolumn}   
\usepackage{bm}        
\usepackage{amssymb}   

\newcommand{\D}{\displaystyle}

\begin{document}


\widetext


\title{Establishing spin-network topologies by repeated projective measurements}
%
\affiliation{Department of Chemical Physics, Weizmann Institute of Science, Rehovot, 76100, Israel}
\affiliation{Fakult\"at Physik, Technische Universit\"at Dortmund, D-44221 Dortmund, Germany}
\author{Christian O. Bretschneider} \affiliation{Department of Chemical Physics, Weizmann Institute of Science, Rehovot, 76100, Israel}
\author{Gonzalo A. \'Alvarez} \affiliation{Fakult\"at Physik, Technische Universit\"at Dortmund, D-44221 Dortmund, Germany}
\author{Gershon Kurizki} \affiliation{Department of Chemical Physics, Weizmann Institute of Science, Rehovot, 76100, Israel}
\author{Lucio Frydman}\email[E-mail address:]{lucio.frydman@weizmann.ac.il}\affiliation{Department of Chemical Physics, Weizmann Institute of Science, Rehovot, 76100, Israel}


\begin{abstract}

It has been recently shown that in quantum systems, the complex time evolution typical of many-bodied coupled networks
can be transformed into a simple, relaxation-like dynamics,
by relying on periodic dephasings of the off-diagonal density matrix
elements. This represents a case of {}``quantum Zeno
effects'', where unitary evolutions are inhibited by projective
measurements. We present here a novel exploitation
of these effects, by showing that a relaxation-like behaviour is endowed to
the polarization transfers occurring within a $N$-spin coupled network. Experimental 
implementations and coupling constant determinations for complex spin-coupling
topologies, are thus demonstrated within the field of liquid-state nuclearmagnetic resonance (NMR).

\end{abstract}

\maketitle

\begin{figure*}
\centering{}
\begin{minipage}[t]{17.8cm}
\begin{minipage}[c]{45mm}
 \begin{center}
\includegraphics[width=46mm,height=55mm]{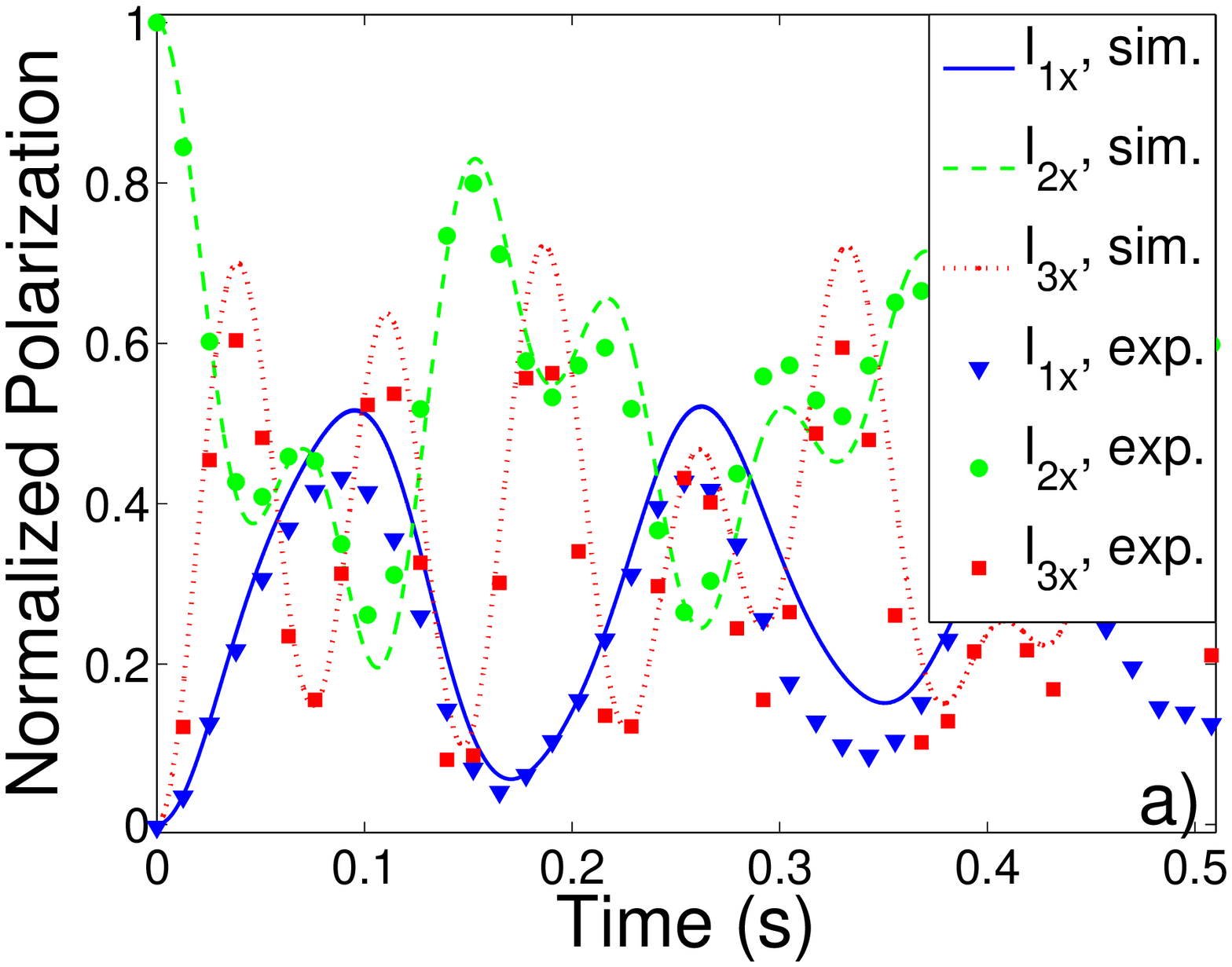} 
\par\end{center}
\end{minipage}
\begin{minipage}[c]{35mm}
\begin{center}
\includegraphics[width=35mm,height=35mm]{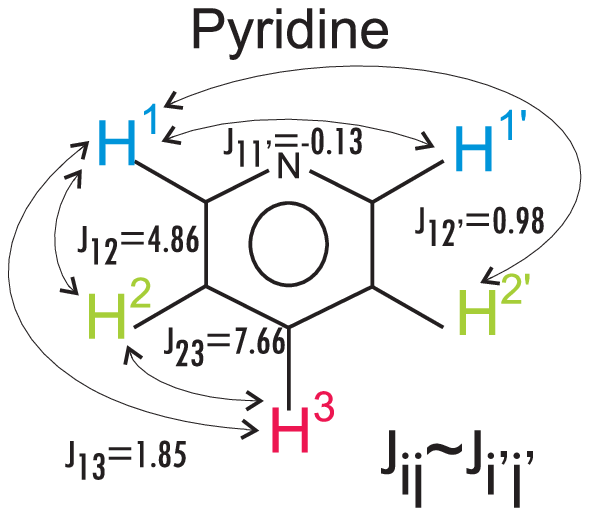} 
\par\end{center}
\end{minipage}
\begin{minipage}[c]{45mm}
\begin{center}
\includegraphics[width=46mm,height=55mm]{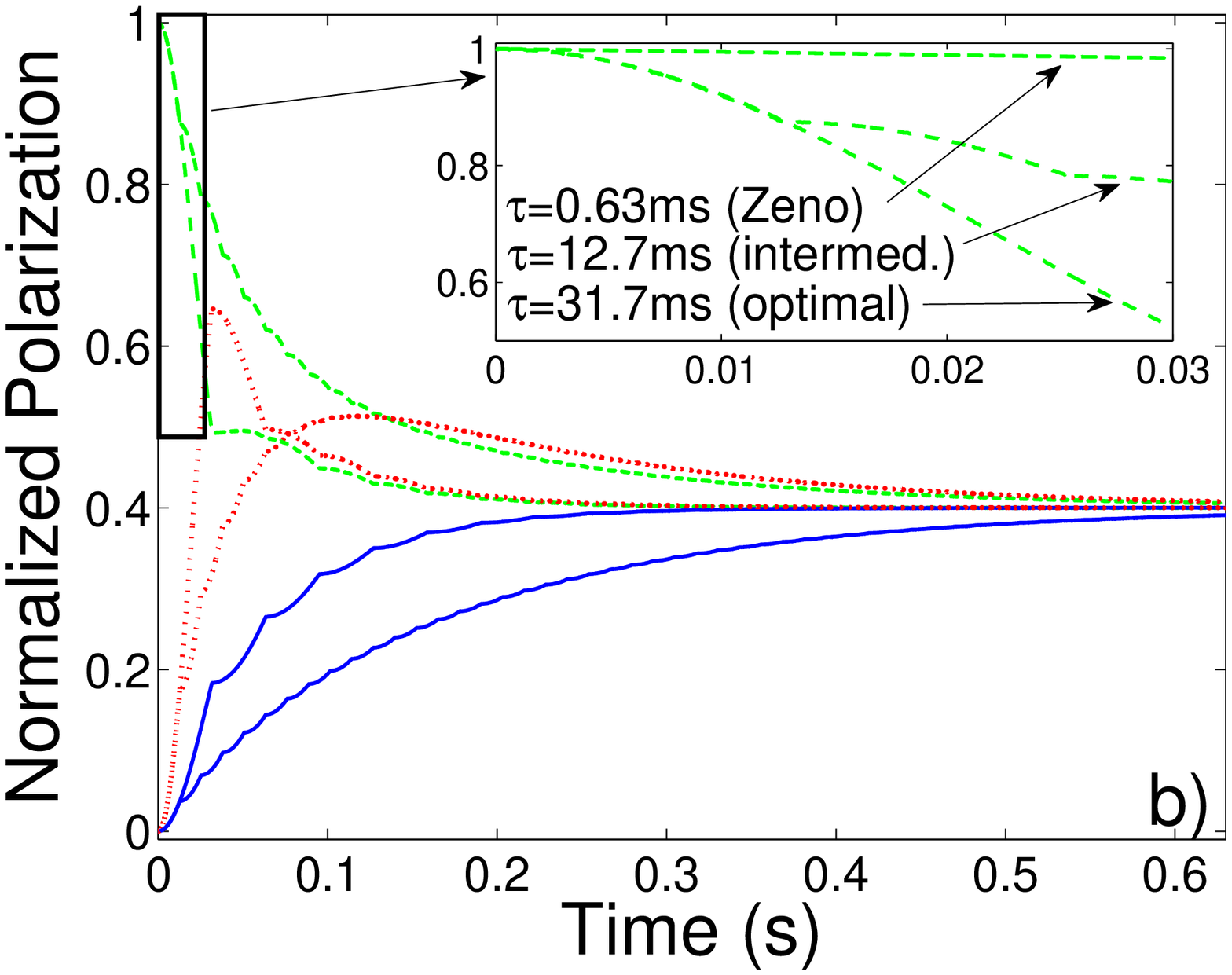} 
\par\end{center}
\end{minipage}
\begin{minipage}[c]{45mm}
\begin{center}
\includegraphics[width=46mm,height=55mm]{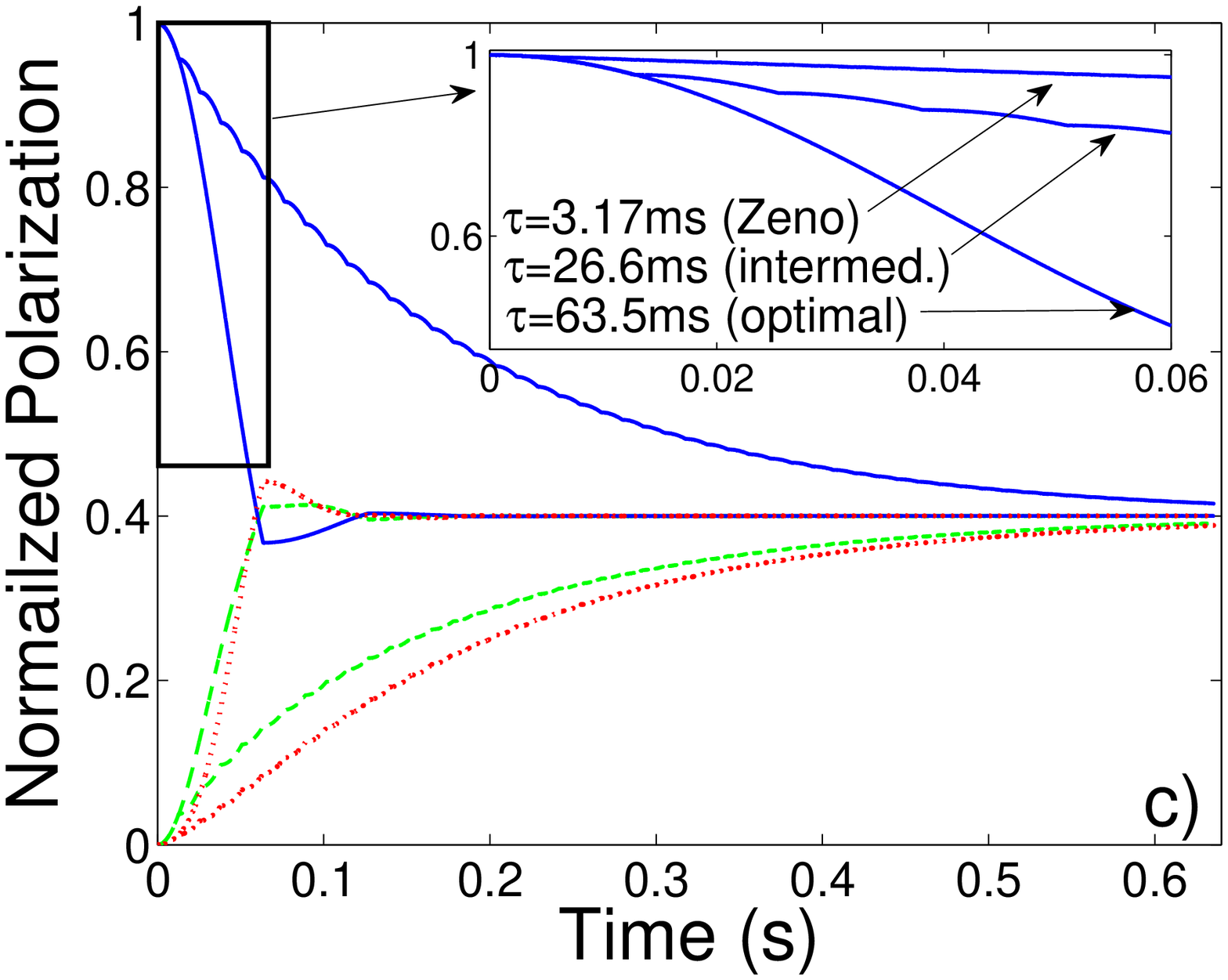} 
\par\end{center}
\end{minipage}
\caption{\label{one} {\small (Color online) Time evolutions of single quantum spin-operators
$\hat{I}_{ix}$ in Pyridine. (a) Coherent oscillatory transfer patterns
(simulated $\rightarrow$ solid/dashed lines, experimental data $\rightarrow$ spin $I_{2}$, blue triangles, spin $I_{2}$, green circles, spin $I_{3}$,  red squares) driven
by $\hat{H}_{J}$ between the spin sites in Pyridine ($J$-couplings
taken from \cite{SDOC}). Implementations involved a
selective preparation of $\hat{I}_{2x}$ (a,b) and $\hat{I}_{1x}$ (c),
followed by the application of a train of $\pi$ pulses
to suppress inter-site chemical shift differences. (b,c) Diagrams
illustrating the switch of the dynamics shown in (a) to quasi-monotonic
polarization transfers, as a result of introducing repeated projective
measurements. These calculated curves involved instantaneous erasements
of the off-diagonal density matrix terms at intervals $\tau$=31.7ms (optimal transfer),
$\tau$=12.7ms (intermed.), $\tau$=0.63ms (Zeno)
for panel (b) and $\tau$=63.5ms (optimal transfer), $\tau$=26.6ms
(intermed.), $\tau$=3.17ms (Zeno) for panel (c). Points in panel (a) (and all remaining data in this work) were
acquired on a 600 MHz NMR spectrometer
using 50 mM Pyridine in CDCl$_3$. The time evolutions of the single quantum spin-operators $\hat{I}_{ix}$ were monitored for
each of the three chemically inequivalent sites in Pyridine in a point-by-point fashion. The transfer pattern under a free evolution
driven by $\hat{H}_J$ was recreated using pulse sequences capable of efficiently suppressing chemical shift differences, including DIPSI-3 \cite{Shak88} and MLEV-8 \cite{Egge92}}}
\end{minipage}
\end{figure*}

{\sl Introduction.}--- A counterintuitive finding in quantum theory concerns the temporal evolution
of systems subject to repeated observations. In the limit of very frequent measurements, this evolution 
can be slowed down almost indefinitely, in what has been nicknamed the quantum Zeno effect \cite{Misr77}. An opposite form can accelerate certain aspects of
unitary evolution of a quantum system, by relying on other forms of projective measurements \cite{Kofm96}.
A particularly suitable field to explore and exploit these Zeno and anti-Zeno quantum regimes, arises in NMR. This connection was first
noted within the framework of a `toy` nutation experiment \cite{Xiao07}. Very recently, 
we have exploited repeated projective measurements in a system composed of spins (qubits) entangled with an effective lattice \cite{Alva10},
to steer polarization transfers between the qubits under mis-matched resonant transfer conditions into either a
quantum Zeno or an anti-Zeno regime. In a series of related studies, Tycko used pseudo-random timings for switching the
coherent character of dipole-dipole dynamics to an incoherent limit \cite{Tyck07}. 

The present study explores the uses of projections within the context of Total Correlation Spectroscopy (TOCSY), a widespread tool in the arsenal of liquid-state
NMR \cite{Sleu96} that serves to establish spin-coupling network topologies. The TOCSY experiment focuses on a homonuclear
network of exchange-coupled spins, whose inequivalent chemical identities (as defined by the action of a chemical shift
term) are made equivalent owing to the application of a
train of suitable, continuous $\pi$ pulses. Under such conditions the effective spin Hamiltonian becomes 
\vspace{-0.25cm}\begin{equation}
\hat{H}_{J}={\displaystyle \sum_{i<j}^{N}J_{ij}\hat{I}_{i}\cdot\hat{I}_{j},}\label{Hj}\end{equation}
leading to complex magnetization transfers modes among all $N$-coupled
spin-${1\over 2}$ partners in a network. As shown in Figs. \ref{one}b, \ref{one}c on further explained {}``projections''
can be used to guide spin evolution to a desired target state
in an almost monotonic fashion. This in turn can facilitate the extraction of information of the $H_J$ coupling structure when dealing with 
complex spin-spin topologies. In general, 
the principles that are here introduced based on repeatedly erasing off-diagonal coherences amount to an effective switch of an $N$-qubit quantum evolution 
from a Hilbert space of dimension $2^{N}$, to a simpler, incoherent form occurring in an $N$-dimensional space.

{\sl Principles and Methods.}--- In the following we assume an $N$ spin-system, subject to the Hamiltonian in Eq. \ref{Hj}.
Even when containing a small number of coupling constants $\left\{ J_{ij}\right\} _{1\le i,j\le N}$
such $\hat{H}_{J}$ can generate a complex time evolution, owing
to its many-body nature \cite{Sleu96}. In fact, the multi-spin response
originated by the Hamiltonian $\hat{H}_{J}$ can be exploited for transferring
coherences among all homonuclear spins within a contiguous structure, and thereby help to establish spin-spin connectivities. Figure \ref{one}a illustrates the complex kind of spin transfer patterns
that can be promoted by $\hat{H}_{J}$. This complexity can eventually become a drawback, as it may demand a systematic scanning of evolution times to ensure that no spin-spin correlations are fortuitously missed. It may also complicate the quantitative determination of the pairwise spin-spin coupling parameters. As illustrated by calculations (Figs. \ref{one}b, \ref{one}c), projective measurements could remedy these disadvantages, and transform the complex evolution imposed 
by $\hat{H}_{J}$ into a quasi-monotonic transfer. The central parameter defining the nature of this transition to a Zeno-like relaxation dynamics is the time delay $\tau$ between projective measurements. As this isotropic $J$-evolution period is reduced, the transfer patterns among the
spins change their originally complex oscillations to a smooth and
almost monotonic polarization sharing. As $\tau$ becomes very short, this turns into a {}``freezing''
(apart from relaxation effects) of all evolutions. It is worthwhile
noting that for optimally chosen $\tau$ values nearly equal distributions
of polarizations - as usually desired in TOCSY - can be rapidly achieved.
In the absence of instrumental losses these amplitudes converge to
the ratio of initially polarized spins and the total number of spins
determined by the quasi-Boltzmann equilibrium \cite{Sake98};
in Fig. \ref{one} this is $\frac{2}{5}$=0.4.

\begin{figure*}
\centering{}
\begin{minipage}[t]{17.8cm}
\begin{minipage}[c]{58mm} 
\begin{center}
\includegraphics[width=60mm,height=30mm]{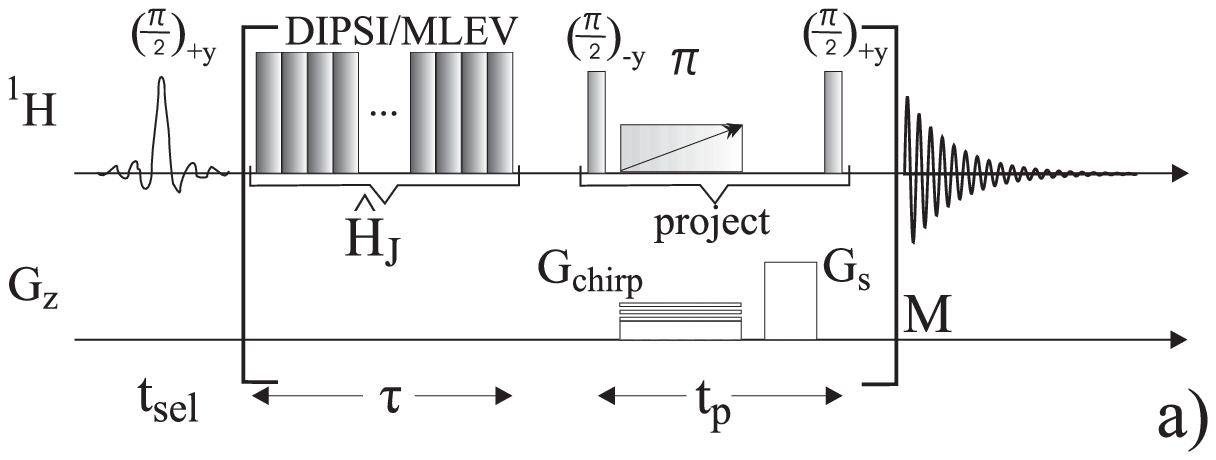} 
\par\end{center}
\end{minipage}
\begin{minipage}[c]{58mm}
 \begin{center}
\includegraphics[width=59mm,height=52mm]{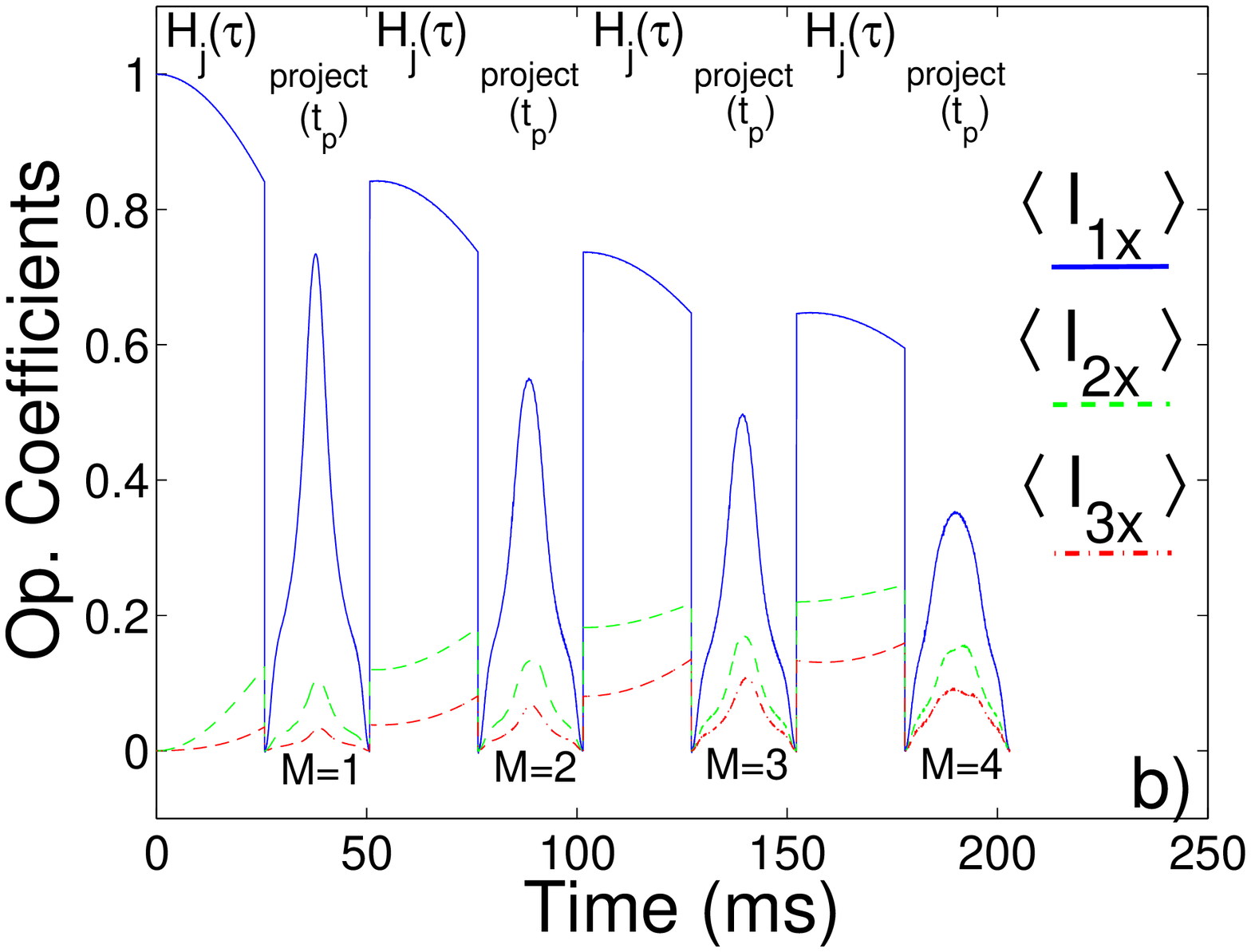} 
\par\end{center}
\end{minipage}
\begin{minipage}[c]{58mm}
 \begin{center}
\includegraphics[width=59mm,height=52mm]{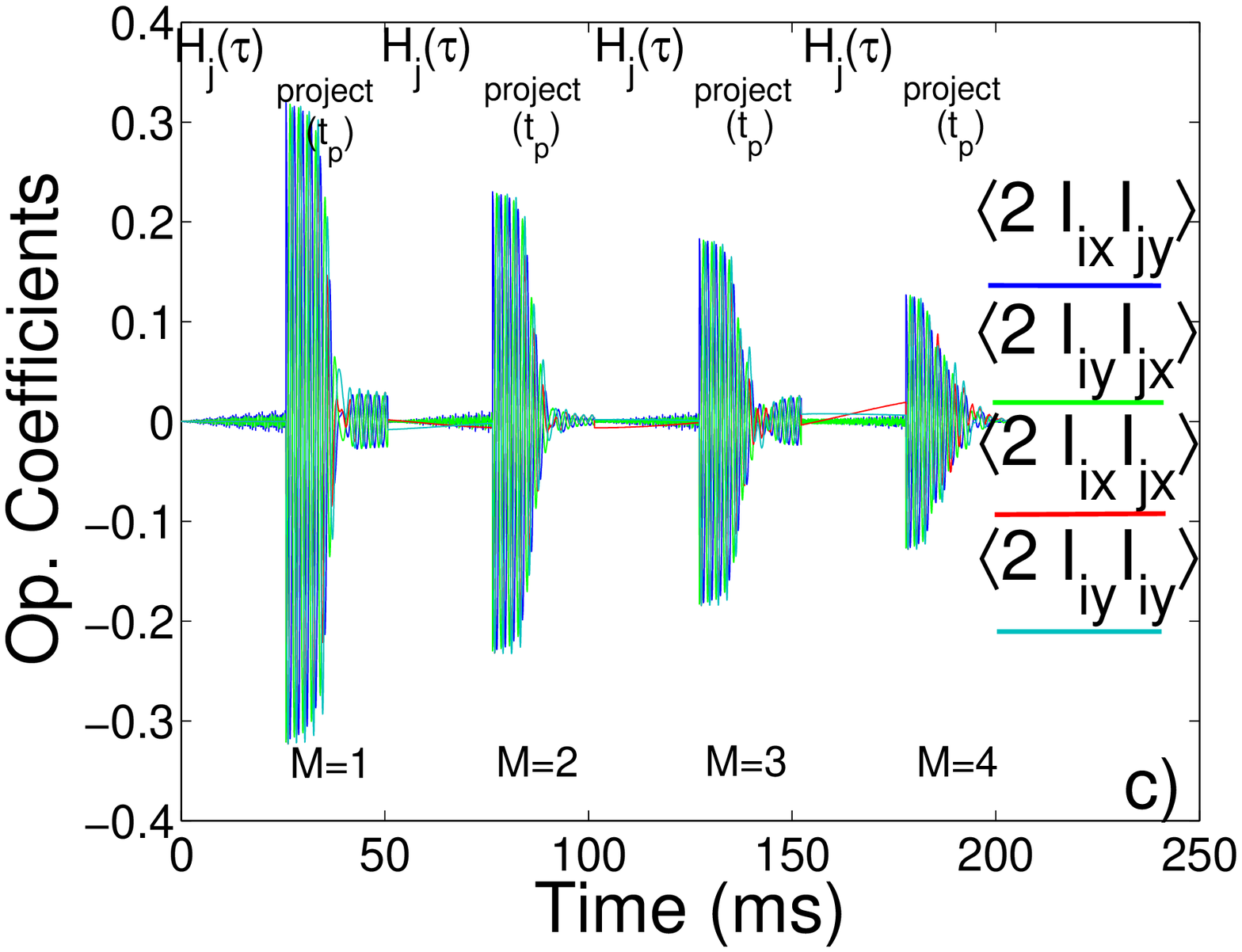} 
\par\end{center}
\end{minipage}
\caption{\label{two} {\small (Color online) (a) Polarization transfer
sequence imitating projective measurements interspersed with segments
of isotropic $\hat{H}_{J}$ evolution. Commonly used parameters were
M=20-40, $t_{sel}$=5-10ms, $\tau$=1-60ms, $t_{p}$=25-30ms,
$G_{s}$=20 G/cm, $G_{chirp}$=0.5-1 G/cm and a sweep range of the
chirped $\pi$ pulses of $\Delta\omega$=20kHz. (b,c) Transfer characteristics 
expected during the course of this sequence for one-spin
single quantum (b) and two-spin zero
and double quantum spin-operators (c). Notice the predicted suppressions
of off-diagonal elements of the density matrix over the full sample
volume, and the steering of the $\langle\hat{I}_{ix}\rangle$ transfer
patterns towards a monotonic evolution.}}
\end{minipage}
\end{figure*}

Figure \ref{two}a sketches how isotropic mixing sequences can be
modified to incorporate the projective measurements leading to the behaviours in Figs. \ref{one}b, \ref{one}c. In lieu
of a full 2D acquisition where every spin is excited and labeled to follow its dynamics,
sequences started with a selective pulse followed by a gradient which either excite or deplete the signal
contributed by an individual spin/chemical site. This continues with a
looped evolution incorporating an isotropic $\hat{H}_{J}$ action
of length $\tau$, a storage of the resulting transverse $\hat{I}_{ix}$
coherences along the z axis of the Bloch sphere, a gradient-based
purging segment ($t_{p}$), and a recall of the stored polarizations
for further evolutions. This {}``mixing'' section of the sequence
was looped $M$ times. Central in this sequence is the purging segment charged
with the erasing of all off-diagonal density matrix elements. This consists of a chirped $\pi$ pulse acting
in combination with a gradient $G_{chirp}$ for dephasing zero-quantum elements \cite{Thri03}, 
followed by a spoil gradient $G_{s}$. The amplitudes of these gradients were varied
in a random fashion throughout each loop in order to avoid fortuitous
echoes of the coherences \cite{Alva10}. 
It is illustrative to examine how the gradient-based elements in Fig.
\ref{two}a help to erase the off-diagonal coherences created by the action of $\hat{H}_{J}$. It can be shown
that in addition to single-spin coherences, this isotropic
Hamiltonian will lead predominantly to two-spin elements involving zero quantum
($I_{i\pm}I_{j\mp}$), single quantum ($I_{iz}I_{j+}$) and double
quantum ($I_{i\pm}I_{j\pm}$) spin-operators \cite{Sleu96}. Even after a $\frac{\pi}{2}$
storage pulse the higher-quantum elements in this series will possess
an off-diagonal character, enabling their cancellation over the full sample
volume when acted upon by the $G_{s}$ gradient (Fig. \ref{two}b). Such dephasing will not happen for
the zero quantum terms, yet these will disappear by the action of the
frequency-swept pulse applied in combination with a random field gradient (Fig. \ref{two}c). 

\begin{figure*}
\centering{}
\begin{minipage}[t]{17.8cm}
\begin{minipage}[c]{5.8cm}
\begin{center}
\includegraphics[width=59mm,height=52mm]{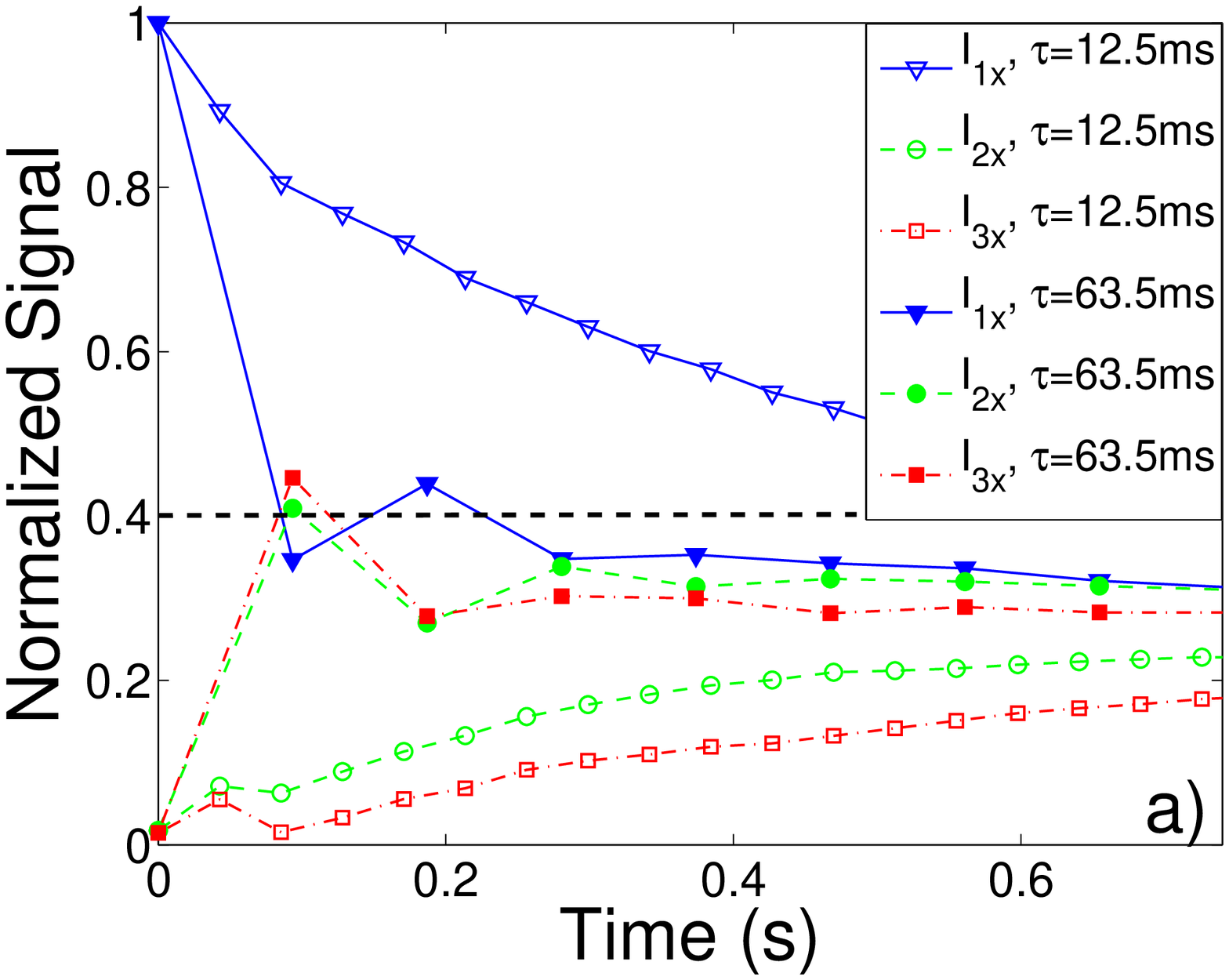} 
\par\end{center}
\end{minipage}
\begin{minipage}[c]{5.8cm}
\begin{center}
\includegraphics[width=59mm,height=52mm]{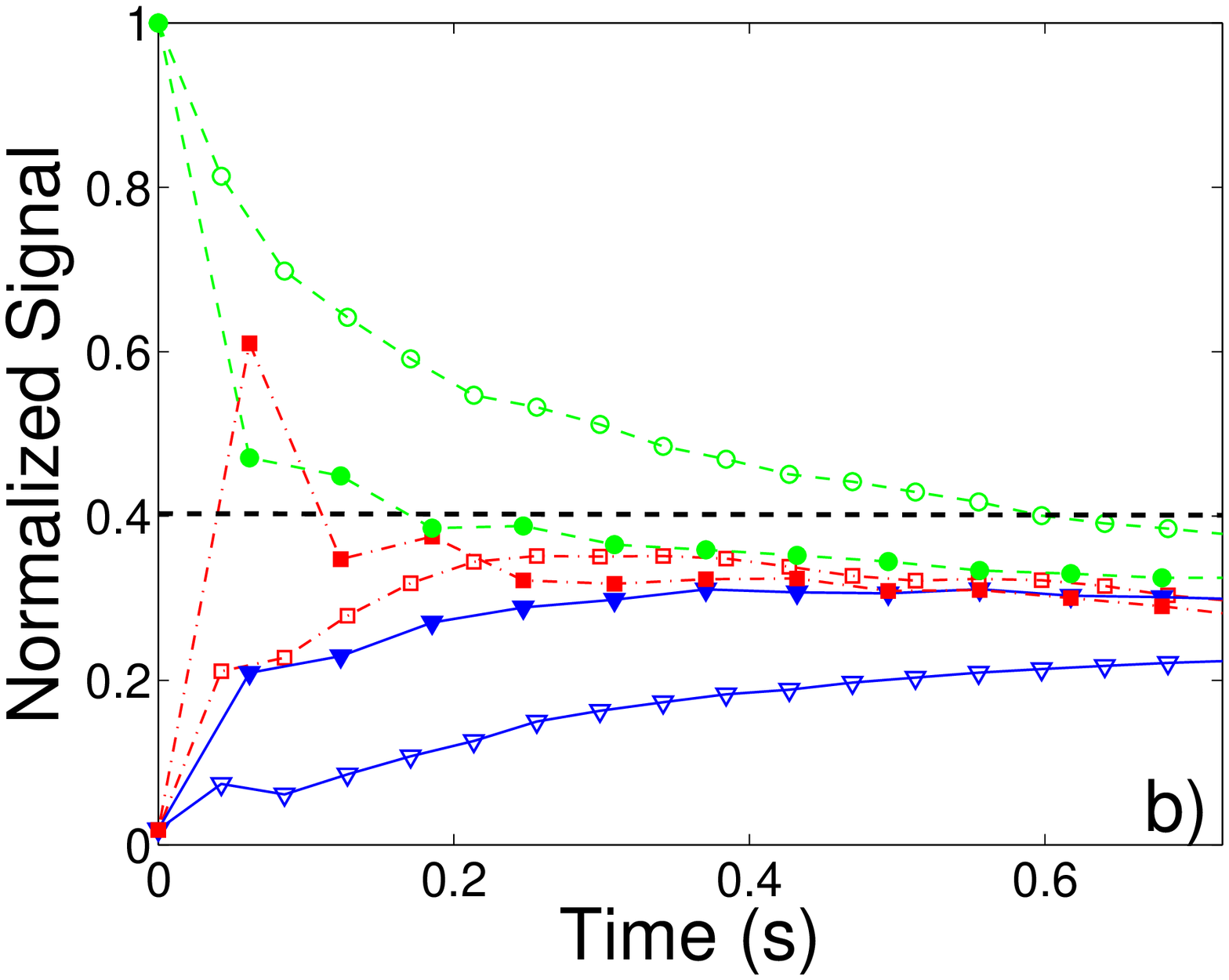} 
\par\end{center}
\end{minipage}
\begin{minipage}[c]{5.8cm}
 \begin{center}
\includegraphics[width=59mm,height=52mm]{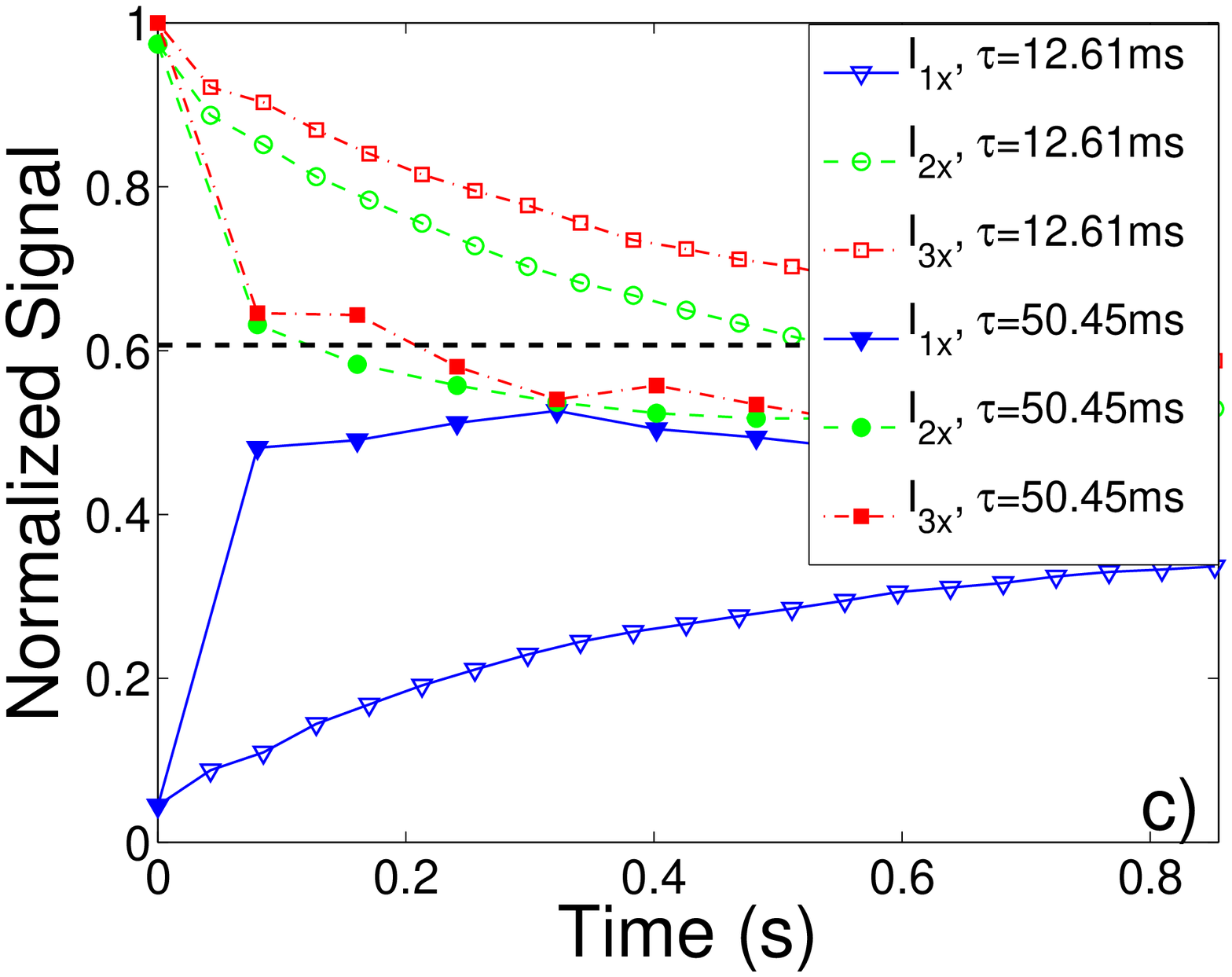} 
\par\end{center}
\end{minipage}
\caption{\label{three} {\small (Color online) Experimental single spin polarizations, 
obtained from the sequence
in figure \ref{two}a for Pyridine. (a,b) Transfer characteristics
resulting from a {}``projected'' $\hat{H}_J$ evolution, following selective
preparations of the spins $\hat{I}_{1}$ (a) and $\hat{I}_{2}$ (b)
for $\tau$=63.5ms (filled symbols) and $\tau$=12.5ms (empty symbols).
(c) Time evolution for an analogous repolarization experiment performed
after a selective demagnetization of spin $\hat{I}_{1}$ for $\tau$=50.45ms
(filled symbols) and $\tau$=12.61ms (empty symbols).}}
\end{minipage}
\end{figure*}

With this sequence at hand, the experimental implementation of the
{}``projected'' $\hat{H}_{J}$ evolution was assayed on Pyridine's protons
for a number of conditions. A series of experimental data are displayed
in Figs. \ref{three}a and \ref{three}b, showing the redistribution
of polarization following the selective excitation of sites $\hat{I}_{1}$/$\hat{I}_{1\prime}$
and $\hat{I}_{2}$/$\hat{I}_{2\prime}$, respectively. The behaviour
of each site's magnetization is fundamentally different from that
shown in Fig. \ref{one}a under the action of $\hat{H}_{J}$, but agrees  
well with the expectations presented in Figs. \ref{one}b, \ref{one}c. 
The main discrepancies between the experimental and the simulated
data arise from the fact that the asymptotic polarization distribution
is reached at a lower signal amplitude ($\sim$0.3) than ideally expected (0.4);
rf pulse inaccuracies and/or relaxation-derived losses are probably responsible for this. 
A related switching to a relaxation-like evolution can be observed if the selective excitation
of a particular site, is replaced by its selective demagnetization.
The corresponding experimental data (Fig. \ref{three}c), illustrates the possibility of repolarizing
a depleted magnetization via neighbouring $J$-coupled spins. For the Pyridine system assayed, the 
theoretical signal amplitudes that should be achieved in such equalization experiments are given by $\frac{3}{5}$=0.6 - very close to the experimentally obtained values.

It follows that projective processes can transform
a complex dynamics like that driven by $\hat{H}_{J}$, into a
quasi-monotonic behaviour. It is worth exploring to what extent can the
coupling constants effecting these transfers, be extracted from such
relaxation-like dynamics. This is facilitated by the short-$\tau$ regime leading to the kind of curves illustrated in Fig. \ref{three}, 
where the original $2^N$ dimensionality defining the coupled spins' wavefunctions is projected into an $N$-dimensional 
space. In such a space only single spin polarizations do matter. The corresponding spin
dynamics can then be represented by a polarization vector $\vec{P}(M\:\tau)$
\begin{equation}\label{math2}
\vec{P}(M\:\tau)=[\boldsymbol{p}(\tau)]^{M}\vec{P}(0). \end{equation}
Here $M$ depicts the number and $\tau$ the timing of the
projective measurements. The vector $\vec{P}$(0) is the initial polarization
of each spin $i$, and the polarization transfer matrix $[\boldsymbol{p}(\tau)]^{M}$ represents an $N\!\times\! N$, 
two-dimensional spectrum whose elements carry the intensities of the self- and of the cross-correlations observed at times $M\cdot\tau$. These
intensities can be simply computed from the knowledge that under the
action of $\hat{H}_{J}$=$J\hat{I}_{i}\cdot\hat{I}_{j}$, the two-spin evolution is of the form \cite{Sleu96}
\begin{equation}
\hat{I}_{iz}\stackrel{\hat{H}_{J}}{\overrightarrow{\hspace{1cm}}}\hat{I}_{iz}\left(\D\frac{1+\cos\frac{J\tau}{2}}{2}\right)+\hat{I}_{jz}\left(\D\frac{1-\cos\frac{J\tau}{2}}{2}\right)+{\cal O}\label{math1}\end{equation}
where ${\cal O}$ represents higher-order relayed transfer terms. Considering short durations $\tau$ and the fact that
projective measurements will only preserve diagonal terms, justify the approximations 
\begin{equation} \D\frac{1+\cos\frac{J\tau}{2}}{2}\!\approx\! 1-\frac{J^2\tau^2}{8} ;\:\:\: \D\frac{1-\cos\frac{J\tau}{2}}{2} \approx \frac{J^2\tau^2}{8}.  \end{equation}
This leads, after $M$ projections, to a two-spin polarization matrix
\begin{equation}
[\boldsymbol{p}(\tau)]^{M}\approx\left[\begin{array}{cc}
1-\frac{1}{2}M\left(\frac{J\tau}{2}\right)^{2} & \frac{1}{2}M\left(\frac{J\tau}{2}\right)^{2}\\
\frac{1}{2}M\left(\frac{J\tau}{2}\right)^{2} & 1-\frac{1}{2}M\left(\frac{J\tau}{2}\right)^{2}\end{array}\right].\end{equation}
The simple quadratic behaviour evidenced by the off-diagonal build-up
in this matrix, can be expanded to the many-body Hamiltonian of Eq. 
\ref{Hj}. Assuming then that the condition $\tau\!J_{ij}\!\!\ll$1 still
holds for all pairwise couplings, the on- and off-diagonal elements of $[\boldsymbol{p}(\tau)]^{M}$
can be approximated by
\begin{equation} [\boldsymbol{p}(\tau)]^{M}  \approx  \left\{ \begin{array}{c} M\frac{J_{ij}^{2}\tau^{2}}{8}\:\:\textit{, for}\:\: i\ne j\nonumber \\ 1-\D\sum_{j}M\frac{J_{ij}^{2}\tau^{2}}{8}\:\:\textit{, for}\:\: i=j \\ \end{array}  \right. \end{equation}
This matrix defines the signal build-up of initially unpolarized
spins $i$. Therefore, the absolute values of the involved couplings can be extracted either
from the initial build-up crosspeaks observed in a two-dimensional spectral distribution, from the ratio between peak intensities obtained in 1D build-up transfer curves, or from these crosspeak values normalized by the self-peak intensity.
For Pyridine - an effective five spin-system - this would
mean that $J_{13}$ and an average of $J_{12}$ and $J_{1^{\prime}2}$ could be extracted
from the short-$\tau$ slopes of the data obtained after a selective
excitation of $\hat{I}_{1}$ (Fig. \ref{three}c), etc. Alternatively,
the coupling constants involved can be established from the decaying
signal amplitude of the initially polarized spin species. For the
model system hereby examined the individual coupling constants were
quantified from the ratios of the signals following projective measurements.
The resulting experimental values are summarized in Table \ref{tab1}, and show a good agreement with values found in literature, 
as well as when compared against the predictions of numerical simulations. 

{\sl Conclusion and Outlook.}--- This study outlined a strategy that uses well-known elements from
NMR spectroscopy's toolbox, like field gradients and rf pulses,
to imitate projective measurements. 
\begin{table}
\begin{centering}
\begin{tabular}{|c|c|c|c|} \hline 
& $\:$ \hspace{0.25cm} $J_{12}$, $J_{1^{\prime}2}$ \hspace{0.25cm} $\:$ & $\:$ \hspace{0.25cm} $J_{13}$ \hspace{0.25cm} $\:$ & $\:$ \hspace{0.25cm} $J_{23}$ \hspace{0.25cm} $\:$ \tabularnewline\hline 
experiments  & 3.4$\pm$0.1 & 2.0$\pm$0.2 & 6.6$\pm$0.9  \tabularnewline\hline
literature  & 4.86, 0.98  & 1.85  &  7.66 \tabularnewline\hline 
simulations  & 3.50  & 1.86  & 7.64 \tabularnewline\hline 
\end{tabular}
\par\end{centering}
\caption{\label{tab1} {\small Experimental J-values extracted from short-$\tau$ peak ratios compared against simulated numerical expectations, and literature values\cite{SDOC}.}}
\end{table}
This allowed us to use NMR as a {}``quantum simulator'' for describing novel Zeno-like effects in exchange-coupled spin networks. These concepts
also lead to a new experimental approach with potentially useful applications, whereby complex and {\sl a priori} unknown
polarization transfer patterns are morphed into smooth, monotonic relaxation-like functions. From these
it is straightforward to determine the underlying spin-spin coupling constants, to establish unambiguous
correlations among spin networks, or to redistribute pools of unused polarization among the spins. Future work will 
show extensions of this strategy to additional NMR experiments, including generalizations to
other types of scalar, dipolar and quadrupolar multi-spin dynamics. We have also found this approach useful 
to shorten recycle delays - particularly in heteronuclear polarization transfer experiments \cite{Kupc07}. It is conceivable that 
the use of dynamical decoupling via frequent $\pi$-pulses can bring about analogous results \cite{DDQZE}.

%
This research was supported by the Israel Science Foundation (ISF
447/2009), the EU through ERC Advanced Grant \#246754)
and a FET Open MIDAS project,
a Helen and Kimmel Award for Innovative Investigation, and the generosity
of the Perlman Family Foundation. G.A.A. thanks the Humboldt Foundation
for financial support and the hospitality of Fakult\"at Physik, TU
Dortmund.
%

\end{document}